# DO PERIODICITIES IN EXTINCTION—WITH POSSIBLE ASTRONOMICAL CONNECTIONS-- SURVIVE A REVISION OF THE GEOLOGICAL TIMESCALE?

Adrian L. Melott[1] and Richard K. Bambach[2]

1. Department of Physics and Astronomy, University of Kansas, Lawrence, KS 66045 USA

2. Department of Paleobiology, National Museum of Natural History,

Smithsonian Institution, PO Box 37012, MRC 121, Washington, DC 20013-7012 USA

ABSTRACT

A major revision of the geological timescale was published in 2012. We re-examine our past finding of a 27 Myr periodicity in marine extinction rates by re-assigning dates to the extinction data used previously. We find that the spectral power at this period is somewhat increased, and persists at a narrow bandwidth, which supports our previous contention that the Nemesis hypothesis is untenable as an explanation for the periodicity that was first noted by Raup and Sepkoski in the 1980's. We enumerate a number of problems in a recent study comparing extinction rates with time series models.

1.  INTRODUCTION

In Melott & Bambach (2010), we examined periodicities in extinction over the last 500 million years, and concluded that a signal detected by Raup & Sepkoski (1984) was present in better resolved, more extensive current data, over a longer time period than they had originally claimed, and at a higher level of significance. The claimed period grew from 26 to 27 Myr, and also is observed now to extend over the entire 500 million year interval rather than just the last 250 million years, due to revisions in the geological time scale since the 1980's. The dating we utilized used the 2004 version of the geological time scale (Gradstein et al. 2004). Nemesis was proposed by Davis, Hut, and Muller (1984) and Whitmire and Jackson (1984) based on the idea that a companion star in a wide orbit perturbs the Oort cloud every 26 Myr, causing comets to enter the inner solar system, some of which collide with the Earth potentially causing mass extinctions. In Melott and Bambach (2010) we concluded that the periodicity was too regular to be attributed to the hypothetical "Nemesis" object, according to orbital perturbation calculations done in the 1980's.

In paleontology, because absolute dating is difficult, primary observations are tied to named intervals in the geological record. These intervals are typically demarcated by different fossil species observed. The assignment of an observation to an interval will not typically change once the named interval has been defined. On the other hand, the dating of interval boundaries, which are determined by interpolation from localities where rocks capable of being dated radiometrically are found in interpretable relationship to rocks containing critical fossils, can change as better outcrops are discovered or more reliable laboratory techniques for dating are devised, and dates are revised accordingly.

Gradstein et al. (2012) constitutes the most recent comprehensive revision of the geological time scale, incorporating the improved dates from the past decade. Periodicities found in the fossil record might not survive the revision of the time scale. In this paper we report on the results when we reassign the data used in Melott & Bambach (2010) to the 2012 geological time scale, and compute power spectra. Also, we discuss some recent objections (Feng & Bailer-Jones 2013) to periodicity detection, note some errors they made in handling or interpreting the paleontological data, which probably invalidate their claims, and show an example of the impact of their procedures on the detection of the signal. We discuss this paper in some detail because feel an explanation of how paleontological data ought to be treated in such analyses is crucial for astrophysicists to understand if they wish to work on topics that bring data on the history of life into calculations related to timing on scales related to galactic phenomena. The results summarized here will be presented in much greater detail elsewhere, including discussion of paleontological details and the nature of the causality.

2. PROCEDURES

2.1 Previous results

In this paper we will present the results of power spectral computation of extinction rates derived from the Sepkoski (2002) compendium (hereafter SEP) of genus origination, extinction, and total biodiversity over the last ~500 Myr.

Motivated by the finding of Raup and Sepkoski (1984) of a 26 Myr extinction periodicity in marine families, Melott & Bambach (2010) undertook to examine this question using new data on genera with revised time scales. We analyzed both the Sepkoski genus data and the Paleobiology Database (PBDB) for periodicity of extinction using the 2004 time scale and found that both display a periodicity in extinction at approximately 27 Myr that extends over 470 Myr. The original finding extended back 250 Myr. The expansion from 26 Myr to 27 Myr is not unexpected, as the geological time scale had expanded by about 3%. We also noted that 10 of 19 mass extinctions as defined by Bambach (2006) lie within ±3 Myr of the maxima in the spacing of the 27 Myr periodicity when the timing is extended over the entire Phanerozoic (the last 542 million years), which differs from a random distribution at the p = 0.01 level.

2.2 Modified data

The only change here in the extinction data from Melott and Bambach (2010) is the alteration of the dates of interval boundaries to coincide with the new 2012 geological time scale. Extinction during an interval is assigned to the end of the interval, because not only are major mass extinctions interval-bounding events, but even ordinary extinctions appear to be primarily pulsed and concentrated at the end of intervals—although of course there is some constant background level of extinction (Foote 2005).

In this study examining extinction data using the revised 2012 time scale we also identify periods of marked extinction intensity based on a technique developed in a study of mass extinctions (Bambach 2006). It was necessary to re-evaluate what intervals were characterized as of marked extinction intensity rather than just using previously designated mass extinction events because adjustments in the time scale altered interval lengths and thus one of the ways in which rate of extinction is expressed. Now we find 19 such intervals occur over the last 470 million years (with another six occurring in the earlier Cambrian Period, for which time correlations are uncertain and evolutionary dynamics are different, and so are not considered here). 10 of the 19 (coincidentally the same numbers as reported in Melott & Bambach 2010 for the 540 million years including the Cambrian Period, but the revised data used in this new study are confined only to the last 470 Myr when time correlation of interval boundaries are secure) lie within ±3 Myr of the maxima in the spacing of the 27 Myr periodicity, which differs from a random distribution at the p = 0.01 level. Although this statistical test uses different criteria, it is dependent on the SEP data, so it is not a fully independent test.

However, the statistical test based on the PBDB data is fully independent, and constitutes a specific check of the hypothesis. One might argue that it is not independent because the data were taken on the same planet. However, the data are predominantly different observations of fossil occurrences, and were handled very differently, with substantial corrections applied, which were not applied to the SEP data.

2.3 Data analysis and results

Computations shown here, as in Melott & Bambach (2010) were performed on AutoSignal 1.7, sometimes supplemented by our own software. We have used both Lomb Periodograms of unmodified data and Fast Fourier Transform of data interpolated (linearly at 1 Myr intervals between adjacent data points). Extensive discussion of the interpolation procedure is found in Melott and Bambach (2011a). Judicious interpolation and ordinary Fast Fourier Transform methods give essentially equivalent results for this data to Lomb-Scargle methods over the range of interesting frequencies. As the interpolation required for FFT constitutes an effective small amount of smoothing, the amplitude of the peak at 27 Myr is slightly reduced in the FFT computation, but even so it reaches a confidence level of p~0.02 assigned by AutoSignal (Figure 1A). This amplitude is higher than found in Melott & Bambach (2010) for the SEP data. Therefore the improvement in the timescale has strengthened the signal. Intuitively, it would seem that a signal which improves with better calibration is a positive sign of its reality.

In Figure 1B we show a Lomb Periodogram (Scargle 1982, 1989) of the PBDB data extinction rates as given by Alroy (2008). Interpolation is not used here for the PBDB data, as its intervals are quite large compared with the 1 Myr steps used in interpolating required for FFT. The Lomb-Scargle procedure (also discussed and compared with FFT in Melott and Bambach 2011a) is essentially a method of constructing a Fourier series for data sampled at irregular intervals—a common occurrence in both paleontology and astronomy. The amplitude shown in Figure 1B is essentially the same as found for PBDB and the 2004 timescale used in Melott & Bambach (2010)—a natural consequence of the fact that the corrections in going from the 2004 to 2012 timescale are relatively smaller compared to the coarse time resolution of PBDB. So, a peak at the same frequency is present in these two independent datasets.

In work to be presented elsewhere (Melott & Bambach 2013) we show a more extensive analysis of both the SEP and PBDB data on extinction, as well as the relationship of the timing of intervals of marked extinction intensity to the phase of the periodicity, and examine the possible causal relationships. This note is restricted to the simple point that the periodicity has survived a rescaling of geological time, and with it, our objections (Melott & Bambach 2010) to the Nemesis hypothesis as a causal mechanism. With the rescaling the periodicity in the SEP data has gone from 27.13 Myr to 27.21 Myr; in the PBDB data it has gone from 26.79 Myr to 26.96 Myr.

3. OBJECTIONS AND THE NEED TO TREAT PALEONTOLOGICAL DATA CORRECTLY

3.1 Recent studies

As this manuscript was being prepared, a paper appeared (Feng & Bailer-Jones 2013, hereafter FB-J) which discussed fits to galactic motion and simple periodicities as compared with "biodiversity" data. They argue that Bayesian statistical analysis does not give any support for either periodicity or coincidence with astronomical, particularly galactic, motion of the Earth. There appear to be a number of problems with this study. We will discuss only those related to the detection of simple periodicity, possibly with an underlying trend. Although some of our comments will apply also to the analysis of relationship to galactic motion, this is a side issue and not treated here. We have already published on how uncertainty and complexity in the galactic mass distribution and velocity field confound precise comparison with terrestrial characteristics (Overholt et al. 2009). We also comment on some related issues related to the detection of the 62 Myr periodicity in marine biodiversity (Rohde & Muller 2005; Lieberman and Melott 2007; Melott & Bambach 2011a,b and references therein), which FB-J claim to re-examine.

Melott & Bambach (2010) cited here was not cited in the FB-J paper. FB-J cite a "Melott and Bambach 2010", but their listed citation was actually published in 2011, and is 2011b here. The "Melott & Thomas 2008" they cite was published in 2009. All of these were already published when their paper was submitted.

3.2     Extinction versus biodiversity as a dependent variable

FB-J discuss the 62 Myr periodicity in marine biodiversity first uncovered by Rohde and Muller (2005). Biodiversity is a product of both origination and extinction. Biodiversity increases when origination exceeds extinction and decreases when extinction exceeds origination. The behavior of just one of these phenomena does not, alone determine biodiversity. Rohde & Muller (2005) and Melott & Bambach (2011a, 2011b) dealt with biodiversity data (the number of kinds of organisms present) when documenting the 62 Myr periodicity in marine biodiversity. However, in their entire paper FB-J only examine extinction rates and events and apparently did not deal with biodiversity (even though they used the term in implying their study showed no strong 62 Myr periodicity), whereas this periodicity in biodiversity has been shown to result from the coherent interaction of extinction and origination fluctuations, and neither origination or extinction alone produce a strong signal (Lieberman & Melott 2007; Melott & Bambach, 2011b). FB-J incorrectly refer to the extinction data as "biodiversity".

3.3     Accumulation of long-lived genera

In assessing the evidence for biodiversity fluctuations, FB-J ignore the substantial effect found by Melott & Bambach (2011b), wherein the effect of the 62 Myr biodiversity periodicity is found to be diluted in the last 150 Myr by the accumulation of long-lived genera, apparently resistant to the cycles. When long-lived genera are removed, the intensity remains strong up to the Recent. This is not a severe impediment to the detection of extinction periodicity, but it would be if they were actually examining biodiversity, as claimed.

3.4     Use of outdated geological timescale

FB-J examine the fractional extinction rate, inferred from the Sepkoski data as reported by Bambach (2006), and inferred from the PBDB data as reported by Alroy (2008). They also examine the timing of mass extinctions also as reported by Bambach

(2006). All these sources of data are based on the 2004 geological time scale. In doing so, there is no indication in the text that they have converted to the 2012 geological time scale, as we have here. The results of our re-analysis of the extinction data using the revised time scale, as reported above, document that the 27 Myr periodicity has somewhat increased strength. As will be seen below, this signal is only significant when the actual timing of most extinction is assigned correctly.

3.5     Formulation of the hypothesis

FB-J importantly describe the hypothesis as, for example, "a periodic model". The Bayes factor (likelihood ratio) (see, e.g. Wang 2010; D'Agostini 2003) is taken as an average computed over all possible phases and periods 10-100 Myr. The substantial past literature on this question does not ask whether "a periodic model" is the best description of the data. In fact, there are theorems, now at textbook level, which show that virtually any mathematical function may be represented as a sum of sinusoids. In these, one frequency will nearly always be dominant. Instead, it is relevant to ask how often random data might produce a dominant frequency comparable in amplitude to that found in the data.

It might be asked whether Bayesian, frequentist, or exploratory data analysis is appropriate to this problem. Bayesian and frequentist methods begin with a problem, go to the data, apply a model—then diverge from one another, either to analysis or to the application of a prior distribution. In exploratory data analysis (Tukey 1977) the next step from data is the analysis of the data to suggest a model—which is then subjected to hypothesis testing. The guiding principle is the statement by Tukey (1962): "Far better an approximate answer to the right question, which is often vague, than an exact answer to the wrong question, which can always be made precise." Exploratory work by Raup and Sepkoski (1984) suggested a periodicity in extinction at 26 Myr on the older timescale; that by Rohde and Muller (2005) a periodicity in biodiversity at 62 Myr. Further work should focus on testing these hypotheses.

Rohde & Muller (2005) did a number of Monte Carlo tests in which they randomized the data. They showed that the probability of any periodicity, that is, at any period, arising at the strength equal to or larger than the one they detected, is 0.01 (or less, depending on details of how the data is reordered). This significance level agrees with estimates made using AutoSignal (e.g. Melott & Bambach 2011a).

This is a crucial point. FB-J assert that random data may produce a strong periodicity, so finding some periodicity that fits is not significant. However, they do not ask how likely this is given the data and given the strength of the spectral peak, nor do they mention the tests by Rohde & Muller. Furthermore, they assert that they are testing the viability of periodic models by averaging over all such models. This of course dilutes any benefit of hitting the "correct" model—which is actually rather well-specified.

3.6     Inappropriate assignment of event timing within intervals

FB-J have assigned both mass extinction events and "continuous" extinction rates to the middle of the interval in question. However, as noted in Foote (2005) there is considerable evidence that most extinctions are pulsed and concentrated at the end of intervals and it is certainly the case for most larger extinction events, because these events are used to set interval boundaries. Therefore it is a better approximation to assign extinctions to the end of intervals than to the midpoints. These intervals range from 7 to 18 Myr for PBDB, and 2 to 9 Myr for the SEP data. For most extinction data, the FB-J choice results in a systematic and varying error of a shift to earlier dates, which will degrade the signal. The degradation will be greater for the 27 Myr signal than for the 62 Myr signal, because the errors constitute a greater phase angle. The effect on obscuring the 27 Myr periodicity of extinction when inappropriate location of extinction within intervals is used will be shown with an example below.

3.7     Overestimate of dating uncertainties in the paleontological record

FB-J assign an uncertainty to the extinction data derived from the length of the interval to which they are assigned.  This is almost certainly (see 3.6) an overestimate of the true uncertainty in date assignments, since the uncertainty in geological date assignments varies with their date, but is now typically less than ± 1 Myr. For example, the end-Permian extinction, the most severe known, is now known to have occurred within 200,000 years of a volcanic ash dated at 252.28 ±0.08 million years ago (Shen et al. 2011) and the end-Cretaceous extinction, famous for the extinction of the dinosaurs, occurred within 30,000 years of 66.043 ±0.05 million years ago (Renne et al. 2013). About 2/3 of all the interval boundaries we consider have 95% errors less than 1 Myr, and only four of them have errors exceeding 2 Myr. None exceed 3.2 Myr. (We must allow for the possibility that systematic error is greater than this random error—but even systematic errors, estimated as the difference between 2004 and 2012 time scale dates, exceed 3 Myr only about 30% of the time.) FB-J consider standard deviations in the extinction timing which are various multiples of the standard deviation of the interval. Consequently, nearly all the FB-J assignment of extinctions are outside the 95% confidence interval for the timing, and more than 2/3 exceed the estimated systematic error. When the assumed uncertainty in the extinction timing is a small multiple, their assignment to the interval midpoint will result in a "miss" for the 27 Myr periodicity. When this is a large multiple, any "hit" will have lowered significance in their Bayesian procedure due to large uncertainty in the data.

3.8    No comparison with independent data on extinction

 FB-J compare the evidence for various models, averaging over all sets of parameters. They correctly note that the Bambach (2006) list of extinction events is not independent of the SEP data.  However, they ignore the fact that the SEP and PBDB data were compiled separately and *are* independent data. Thus, one can formulate the hypothesis in a new way: is the biodiversity period found in the SEP data also found in the PBDB data? This question was already answered affirmatively in Melott (2008) and Melott & Bambach (2011a); in the latter the same periodicity was also found in a third data set. The same is true for the 27 Myr periodicity of extinction. It occurs in both SEP and PBDB data, as reported in Melott & Bambach (2010) for the 2004 version of the time scale and in this paper. Two additional statistical tests are possible, as shown those papers, in which the question is formulated as finding agreement of a period with that already found in independent data. FB-J chose not to do such tests.

3.9   No comparison with related geological data
   As related to the question of testing the general question of periodicity versus specific sets of parameters, FB-J also ignore the finding of periodicities coincident with biodiversity fluctuations within the errors in period and phase in the emplacement of sediments (Melott & Bambach 2011b; Meyers and Peters 2011) and strontium isotope ratios in seawater (Melott et al. 2012). Of course, as a formally constituted statistical question, these are not relevant: FB-J were claiming to test for biodiversity signals.  But the practical scientific issue is that these variables are known to be related to biodiversity and extinction, as discussed in these references.

3.10   An example of some effects: timescale and timing assignment
   For all these reasons, we do not accept the conclusions of FB-J. As an example of the effects of some of their errors, we show in Figure 2B the power spectrum of the SEP data extinction rates if we assign our data to the mid-points of the intervals and use the 2004 geological time scale, paralleling the choice made by FB-J. This can be compared directly with Figure 2A, a power spectrum of SEP data using the 2012 geological time scale and assigning the data to the interval end boundaries, where we know they are concentrated. As can be seen, although the 27 Myr peak is still visible in Fig. 2B, its amplitude is degraded. Treating the data as FB-J treat it reduces the significance of the signal that is unambiguously present when the data are treated more correctly. The significance of the spectral peak which is p ~ 0.02 in Fig. 2A, is reduced to p ~ 0.2 in Fig. 2B.

4.   ASSESSMENT USING AKAIKE WEIGHTS
   The method of Akaike weights (Wagenmakers & Farrell 2004) is a popular method for converting the Akaike information criterion (based on the likelihood) into a conditional probability for each of a set of possible models into a probability that additional data would support that model. FB-J present 19 different models for the parent distribution of probabilities from which extinctions are drawn. At the suggestion of the referee, we have applied this method to the likelihood values in FB-J, Table 7, using the numbers of free parameters for these models given in Table 3 and the text. We use the set of extinctions in Bambach (2006), called B18 in FB-J, which is the the largest data set applicable to all of the models. We find that PNB (random draws from a periodic function) has a probability of 89.5% of being the "best" model of the 19. The next-best is RNB (random draws from a uniform background) with a probability of 5.6%. Clearly their analysis, using the maximum likelihood method evaluated as a probability, shows a preference for a periodic parent distribution behind mass extinction events.

## 5. CONCLUSIONS AND DISCUSSION

We conclude that the updates in the 2012 geological time scale somewhat strengthen the signal for a 27 Myr periodicity in extinction, using the Sepkoski (2002) data, which has the best time resolution of any available. The same periodicity is found in the independent PBDB extinction data, as provided by Alroy (2008). Therefore our objections to the Nemesis hypothesis as an explanation for the periodicity, as previously published (Melott & Bambach 2010), stand.

We note a number of problems in the treatment of the paleontological data by FB-J, which reduce their ability to detect any periodicity in either biodiversity or extinction. We also note that in constructing their test which concluded that periodic models were not better than random models, they averaged over periodic models with all frequencies and phases, which via the Central Limit Theorem essentially randomizes the periodic models, rather than testing previously elaborated questions with new fossil data. It appears that the formulation of the question in this way would eliminate the significance of any candidate model. The question can alternatively be formulated as the agreement of independent sets of data. Their maximum likelihood results, interpreted as probabilities via the method of Akaike weights, strongly favor a periodic model for the timing of mass extinction events.

We note how it is necessary to use paleontological data correctly to analyze pattern in the history of life that may relate to phenomena on time scales of galactic phenomena. If we assign extinctions to the midpoints of intervals as they did, and use the older timescale, we also would have concluded that the 27 Myr periodicity was not significant.

The highly regular period and its relatively narrow bandwidth suggest but do not require an astronomical role in terrestrial extinction events. However, the varying nature of these events, as discussed in Melott and Bambach (2013), does not admit of an obvious interpretation.

## 6. ACKNOWLEDGMENTS

We thank Hume Feldman, Bruce Lieberman, Dave Raup, Graham Wilson, and especially the referee Michael Foote for helpful comments on the manuscript. This is Paleobiology Database publication # 181.

Figure Captions

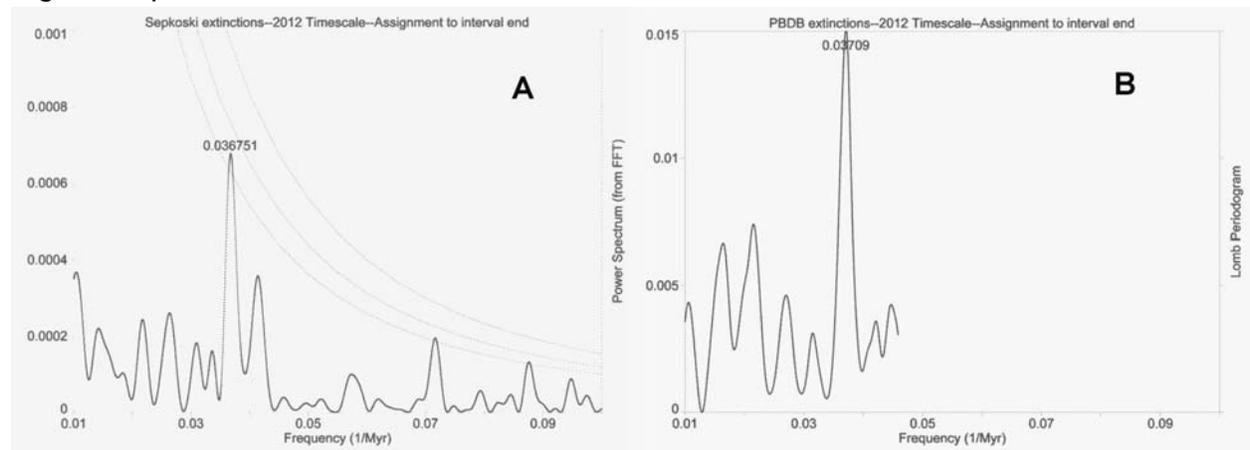

Figure 1 (A) The power spectrum of extinction in the Sepkoski data, computed by FFT. The noted peak is near 27 Myr. The curved lines indicate confidence levels of .05, .01, and .001. (B) A Lomb Periodogram of the Paleobiology Database extinction rate (see text). There is a spectral peak at 27 Myr. The difference in amplitude between the two plots is not significant, as it arises from a different procedure for defining extinction rates as given in the two data sets. The 2012 geological time scale is used and extinction data are assigned dates at the interval end boundaries.

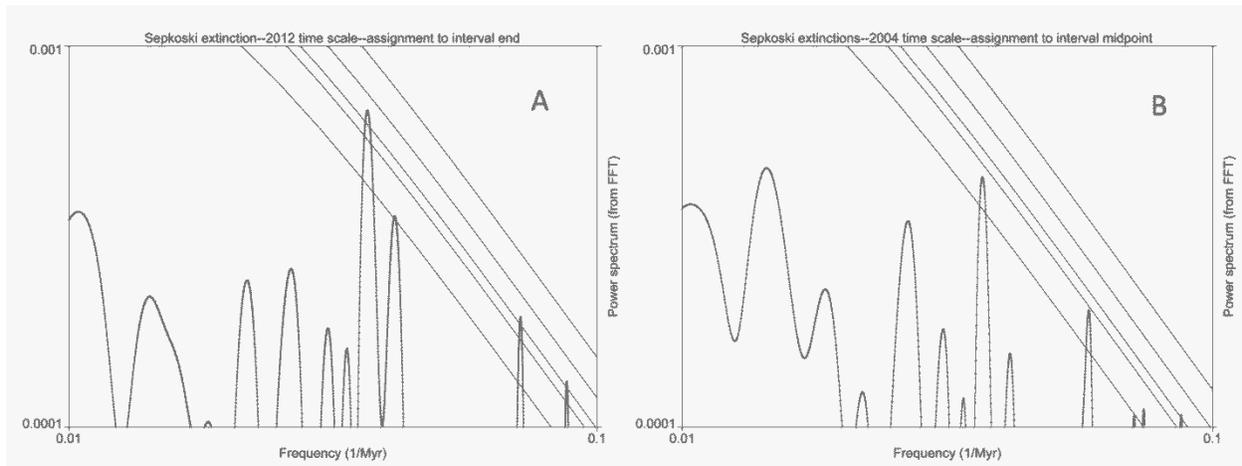

Figure 2. (A) A log-log plot of the power spectrum of extinction in the Sepkoski (2002) biodiversity data using the 2012 geological time scale. Extinction assignments are to the end of given time intervals. The noted spectral peak corresponds to a period of 27.1 Myr. The parallel lines indicate confidence levels of $p$ = 0.5, 0.1, 0.05, 0.01, and 0.001. This can be compared with B to show the effect of timescale and assignment of extinctions to the middle versus end of intervals. (B) Same as A, except that the 2004 geological time scale is used, and data assignments are made to the middle of geological intervals, as in FB-J. This demonstrates the loss of amplitude in the primary signal when making this choice. The confidence level is degraded from $p$ = 0.02 to about 0.2.